\documentclass[letterpaper, 10 pt, conference]{ieeeconf}  

\IEEEoverridecommandlockouts                              
\usepackage{cite}
\usepackage{amsmath,amssymb,amsfonts}
\usepackage{algorithm}
\usepackage{algorithmic}

\usepackage{graphicx}
\usepackage{textcomp}
\usepackage{xcolor}

\newtheorem{definition}{Definition}
\newtheorem{proposition}{Proposition}

\title{\LARGE \bf
Estimation of a sparse multi-qubit Hamiltonian via compressed sensing
}

\author{ Juntao Tu\textsuperscript{1}, Yuanlong Wang\textsuperscript{2}, Shuming Cheng\textsuperscript{3}, Shuixin Xiao\textsuperscript{4},
Zhibo Hou\textsuperscript{5}
\thanks{*This work was supported by the Quantum Science and Technology-National Science and Technology Major Project (Grant No. 2023ZD0301400), the National Natural Science Foundation of China (Grant No. 12288201), and the CAS Project for Young Scientists in Basic Research (YSBR-131).}
\thanks{$^{1,2}$State Key Laboratory of Mathematical Sciences, Academy of Mathematics and Systems Science, Chinese Academy of Sciences, Beijing 100190, China, and School of Mathematical Sciences, University of Chinese Academy of Sciences, Beijing 100049, China.
{\tt\small tujuntao@amss.ac.cn, wangyuanlong@amss.ac.cn}}%
\thanks{$^{3}$College of Electronic and Information Engineering, Tongji University, Shanghai 201804, China, Shanghai Research Institute for Intelligent Autonomous Systems, Tongji University, Shanghai 201203, China, and Institute for Advanced Study, Tongji University, Shanghai 200092, China. 
{\tt\small shuming\_cheng@tongji.edu.cn}}
\thanks{$^{4}$Department of Electrical and Electronic Engineering, University of Melbourne, Parkville, VIC 3010, Australia.
{\tt\small shuixin.xiao@unimelb.edu.au}}
\thanks{$^{5}$Laboratory of Quantum Information, University of Science and Technology of China, Hefei 230026, China, CAS Center For Excellence in Quantum Information and Quantum Physics,
University of Science and Technology of China, Hefei 230026, China, Hefei National Laboratory, Hefei 230088, China. 
{\tt\small  houzhibo@ustc.edu.cn}}
}

\begin{document}

\maketitle
\thispagestyle{empty}
\pagestyle{empty}

\begin{abstract}

Hamiltonian estimation is an effective approach in studying the structure and dynamical evolution of quantum systems. The difficulty in estimating the Hamiltonian is that an $N$-qubit Hamiltonian has $4^N-1$ unknown parameters, requiring exponentially many equations for information extraction. In this paper we develop a method based on compressed sensing to estimate the Hamiltonian of a multi-qubit system. We identify a problem where as $N$ increases, the common sufficient condition (Restricted Isometry Property) for compressed sensing often fails, obstructing the application of compressed sensing in ($N\geq 3$)-qubit Hamiltonian estimation. To solve this problem, we propose a ``scale transformation" technique to restore RIP and ensure a compressive estimation of a $k$-sparse Hamiltonian using only $O(k\log(4^N/k))$ equations. In the numerical examples, we estimate the Hamiltonians of two 6- and 30-qubit systems, demonstrating the effectiveness of the method.
\end{abstract}

\section{INTRODUCTION}\label{INT}
In recent years, quantum technology has demonstrated significant potential and  application value across various fields, leading to the emergence of numerous interdisciplinary research fields, such as quantum computing \cite{QuantumComputation}, quantum communication \cite{Quantum-communication}, quantum sensing \cite{Quantum-sensing}, and quantum cryptography \cite{qcode}. A common thread across all these domains of quantum technology involves the utilization and investigation of quantum systems. Consequently, a deeper understanding of quantum systems is fundamental to sustaining the development of quantum technologies \cite{Nielsen_Chuang}.

Hamiltonian, as the generator of the unitary evolution operator of a closed quantum system, is a key physical quantity determining its evolution. Estimating all or the major parameters of  the Hamiltonian is vital for our understanding of the system structure \cite{DONG2022243}, and fundamental for subsequent tasks such as noise suppression \cite{Wang2017} or optimal control of the quantum system \cite{PRXQuantum.2.030315}.

As the qubit number $N$ of an artificial quantum system (especially quantum computers) increases rapidly, the number of parameters in the Hamiltonian grow exponentially as $4^N-1$, bringing huge experiment burden for the full estimation of the system Hamiltonian. For example, Ref. ~\cite{two_step_alg} proposed a two-step optimization algorithm to fully identify an unknown quantum Hamiltonian based on quantum process tomography, where the number of measurement equations required is in fact exponential on the qubit number. When part of the parameters in the Hamiltonian, including their values and positions, are known a priori, the number of equations needed can often be reduced to around the level of the number of unknown parameters, which can be much smaller than $4^N-1$ in special systems. There is much research in this topic, such as Refs. \cite{QHI-Measurement-Time-Traces,PhysRevLett.Resonant.Coupling,Hamiltonian-Identification-via-QEC}.

From the view of classical signal processing, an intermediate scenario between the above two scenarios is when we know a prior the signal to be estimated (denoted as $h\in\mathbb{R}^n$) is sparse, i.e., $h$  includes at most $k$ non-zero elements with their positions unknown and it holds that $k\ll n$. This scenario often holds through prior knowledge or stems from the need to estimate a sparse approximation to an unknown signal, and compressed sensing (CS) method can be potentially an efficient approach. In fact, many signal reconstruction problem can be modeled as solving a linear system equation, and CS theory demonstrates that if the coefficient matrix of this linear system satisfies the Restricted Isometry Property (RIP) \cite{CS-intro} and the solution $h$ is sparse, then $h$ can be accurately estimated using a significantly smaller number of equations that is $O(k\log(n/k)\ll n$. This approach has found wide application in areas such as image \cite{image} and video
compression \cite{video}, medical imaging (e.g., MRI acceleration \cite{MRI}), single-pixel photography \cite{PIX}. More recently, CS theory has also been introduced to solve quantum parameter estimation problems such as phase estimation \cite{QUA}, state tomography \cite{PRL-QPT}, process tomography \cite{PRB1}, noise spectroscopy \cite{PRAPL}, etc. Especially, Ref.~\cite{PRA-Main} introduced the compressed sensing method into Hamiltonian estimation and provides a framework for modeling the problem as a linear sparse estimation problem.

However, when scaling up the qubit number, the CS estimation of the system Hamiltonian still present certain aspects worthy of consideration. We find that as the qubit number increases, the RIP of the measurement matrix might not be naturally satisfied, making the CS method infeasible. As a solution, we propose a method termed ``scale transformation". We demonstrate that if the initial product states and the final observables are randomly chosen according to proper symmetry probability distributions, an equivalent estimation problem can be established where the new unknown vector is a rescaled version of the original unknown vector, while the equivalent problem guarantees the RIP of the measurement matrix, rendering the CS method applicable again. We further develop a second-order correction method to fit the second-order terms in the  measurement data and thus to reduce the approximation error in data. We perform numerical simulations on two 6- and 30-qubit systems to demonstrate the effectiveness of our CS approach and second-order correction method.

The organization of this paper is as follows. Section \ref{PRE}
introduces some preliminary knowledge of quantum dynamical parametrization and formulates this Hamiltonian estimation problem. Section \ref{MR} presents our main result, including compressed sensing algorithm for Hamiltonian estimation, scale transformation, second-order correction method and the establishment of an optimal control problem as an application example based on the Hamiltonian estimation result. Numerical examples are presented in Section \ref{NE}
and Section \ref{CL} concludes this paper.

\section{Problem formulation}\label{PRE}
 
Here, we give a brief introduction to quantum systems and dynamics. Every deterministic quantum state $\vert \psi\rangle$ is a vector in a $d$-dimensional Hilbert space, and the classical probabilistic mixture of multiple deterministic states, i.e., a mixed quantum state, is described by a density matrix $\rho\in\mathbb{C}^{d\times d}$ satisfying $\rho=\rho^{\dagger}$ ($\dagger$ represents conjugate and transpose), $\rho\geq0$ (positive semidefinite) and $\mathrm{Tr}(\rho)=1$. Observables, physical quantities that can be measured, are described by Hermitian matrices in $\mathbb{C}^{d\times d}$. 

In quantum mechanics, measuring a given observable $M$ on a state $\rho$ does not yield a deterministic outcome; rather, the $i$-th eigenvalue $\lambda_i$ of $M$ is obtained with probability $p_i\equiv\mathrm{Tr}(\rho_i\rho)$ which represents the probability of $\rho$ collapsing to $\rho_i$, the eigenvector of $M$ corresponding to $\lambda_i$. The quantum expectation, averaged over all possible outcomes when measuring $M$, is defined as $\langle M\rangle_{\rho}\equiv \mathrm{Tr}(M\rho)$. 

Pauli matrices are
 \begin{equation}
     X=\begin{pmatrix}
         0&1\\
         1&0
     \end{pmatrix},
     Y=\begin{pmatrix}
         0&-\text{i}\\
         \text{i}&0
     \end{pmatrix},
     Z=\begin{pmatrix}
         1&0\\
         0&-1
     \end{pmatrix},
 \end{equation}
and these matrices are also denoted as $\sigma_1,\sigma_2,\sigma_3$. The identity matrix is $I$, which in $\mathbb{C}^{2\times 2}$ is also denoted as $\sigma_0$. For a single-qubit state $\rho$, the Bloch sphere can be constructed using $X, Y, Z, I$ as
\begin{equation}
    \rho=(I+xX+yY+zZ)/2,
\end{equation}
where $(x,y,z)$ could be any point in the unit sphere in $\mathbb{R}^3$.
 \begin{definition}[1, 2-norm of vector]
     For a vector $\vec{a}\in \mathbb{R}^n(\mathbb{C}^n)$, the 1, 2-norm of $\vec{a}$ is $\|\vec{a}\|_1\equiv\sum_{i=1}^n\vert{a_i}\vert$,
     and $\vec{a}$ is $\|\vec{a}\|_2\equiv\sum_{i=1}^n\vert{a_i}\vert^2$,  where ${a_i}$ is the $i$-th component of $\vec{a}$.
 \end{definition}

 The time evolution of a quantum state can be described by the Liouville–von Neumann equation(we set $\hbar=1$ in this paper)
\begin{equation}\label{LN eq}
    \dot{\rho}(t) =-\text{i}[H,\rho(t)],
\end{equation}
where $[A,B]\equiv AB-BA$ is the commutator. In a multi-qubit system, the Hamiltonian description is general enough to allow for multi-axis, complicated and non-classical correlations between different qubits. Theoretically, the estimation of the Hamiltonian $H$ of a system can be performed as follows: for a set of test states $\{\rho_i\}$, measure a set of observables $\{M_j\}$ after a time evolution of duration $t$. This gives a series of equations
\begin{equation}\label{ob eq}
    y_{ij}\equiv\langle M_j\rangle_{\rho_i}\equiv\mathrm{Tr}\left(e^{\text{i}tH}M_je^{-\text{i}tH}\rho_i\right).
\end{equation}
Therefore, \eqref{ob eq} provides us with a set of equations that describe the input-output relationship of the quantum system. With a sufficient amount of observational data, all information about the Hamiltonian parameters can be extracted, which forms the foundation for our parameter estimation. Following the route of \cite{PRA-Main}, we consider the Dyson series expansion and obtain
\begin{equation}\label{dyep}
    \langle M_j\rangle_{\rho_i}=\mathrm{Tr}(M_j\rho_i)+\text{i} t\mathrm{Tr}([H,M_j]\rho_i)+O(S^2t^2),
\end{equation}
where the constant $S$  equals to the spectral norm of the superoperator $-\text{i}[H,\,\cdot\,\,]$. In contrast to \cite{PRA-Main}, here we rewrite \eqref{dyep} in an equivalent form separating the knowns from the unknown, as
\begin{equation}\label{dyeptran}
    \langle M_j\rangle_{\rho_i}=\mathrm{Tr}(M_j\rho_i)+\text{i} t\mathrm{Tr}(H[M_j,\rho_i])+O(S^2t^2).
\end{equation}
Therefore, these equations can be approximated as linear by discarding the terms with orders higher than one, when the evolution time  $t\ll S^{-1}$.

Next, we consider the parametrization of the Hamiltonian. For an $N$-qubit system, we take an orthonormal basis $\{\Gamma_\alpha\}$ with respect to the inner product $\langle A,B\rangle\equiv \text{Tr}(A^\dagger B)$, with each $\Gamma_\alpha\in\mathbb{C}^{2^N\times 2^N}$ Hermitian. We can expand $H$ as $H=\sum_{\alpha}h_{\alpha}\Gamma_{\alpha}$. In the language of frame theory [8], $\{\Gamma_\alpha\}$ constitutes a Parseval frame of the underlying Hilbert space of $H$. In particular, to adapt to the common physics background, we choose a special choice of this frame, where $\{\Gamma_\alpha\}$ are all the possible tensor products of Pauli matrices and $I_{2\times 2}$. We substitute the orthonormal expansion into \eqref{dyep} and neglect the high-order term to reach.
\begin{equation}\label{linar eq}
    y_{ij}=\langle M_j\rangle_{\rho_i}=\mathrm{Tr}(M_j\rho_i)+\text{i}t\sum_{\alpha}\mathrm{Tr}(\Gamma_{\alpha}[M_j,\rho_i])h_{\alpha}.
\end{equation}
We denote
\begin{align}
    \bar{y}_{ij}\equiv&y_{ij}-\mathrm{Tr}(M_j\rho_i),\\
    \nonumber \phi_{ij}\equiv&\text{i}t(\mathrm{Tr}(\Gamma_{0}[M_j,\rho_i]),\mathrm{Tr}(\Gamma_{1}[M_j,\rho_i]),\cdots,\\
    &\mathrm{Tr}(\Gamma_{{4^N-1}}[M_j,\rho_i]))^T,\\
    h\equiv&(h_{0},h_{1},\cdots,h_{{4^N-1}})^T,
\end{align}
and then \eqref{linar eq} is transformed to
\begin{equation}\label{dot eq}
    \bar{y}_{ij}=\phi_{ij}^T h.
\end{equation}
We take $m$ pairs $(M_j,\rho_i)$ randomly according to a proper probability distribution, and stack the $m$ vectors $\phi_{ij}^T$ into an $m\times4^N$ matrix $\Phi$. Then by stacking $\bar{y}_{ij}$ into a vector $\bar{y}$, we have the final linear system of equation
\begin{equation}\label{mat eq}
    \bar{y}=\Phi h.
\end{equation}
Then one performs $m$ experiments according to the setting of \eqref{mat eq}, and processes the measurement data to obtain the approximate value of the LHS of \eqref{mat eq}. By further performing a proper CS algorithm, one can estimate $h$, and finally infer the Hamiltonian.

\section{Main result}\label{MR}

\subsection{CS on Hamiltonian Estimation of Multi-qubit System }\label{CSAL}
First, we briefly introduce the theory of compressed sensing. CS is a signal processing method that allows for the efficient acquisition and reconstruction of signals from far fewer samples than required by the classical Nyquist–Shannon sampling theorem, provided that the signal is sparse or compressible in some known basis. The core idea is that if a sparse signal is sampled randomly under proper conditions, it can be recovered correctly with a high probability. 

To apply CS to practical Hamiltonian estimation problems, it is first necessary to model the problem in an appropriate form, a task already accomplished in Section \ref{PRE}. However, successfully completing the estimation also requires satisfying several conditions. We first examine the measurement matrix $\Phi$ in \eqref{mat eq}, which is an $m \times 4^N$ matrix. A standard linear inversion solution of \eqref{mat eq} requires $m\geq 4^N$ experiments. Nevertheless, CS allows for a unique solution to \eqref{mat eq} even when $m\ll4^N$, provided that the measurement matrix satisfies the RIP and the solution vector satisfies the sparsity condition. 
\begin{definition}[$k$-sparse]
    A vector $\hat{h}$ is $k$-sparse if it has at most $k$ nonzero elements. 
\end{definition}
\begin{definition}[$k$-sparsifying]
    To $k$-sparsify a vector means retaining the $k$ largest elements of the vector and setting all the remaining elements to zero.
\end{definition}    
\begin{definition}[RIP]
    The measurement matrix $\Phi$ satisfies RIP if and only if with a minimal given constant $\delta_k\in(0,1)$, for with any two $k$-sparse vectors $\hat{h}_1$, $\hat{h}_2$, it holds
        \begin{equation}\label{RIP}
            \begin{split}
                (1-\delta_k)\|\hat{h}_1-\hat{h}_2\|_2\leq&\|\Phi(\hat{h}_1-\hat{h}_2)\|_2\\
                \leq&(1+\delta_k)\|\hat{h}_1-\hat{h}_2\|_2.
            \end{split}
        \end{equation}
\end{definition}
In this paper we assume to know a prior that the Hamiltonian vector $h$ to be estimated is $k$-sparse. This assumption can be taken in proper physical systems with enough prior knowledge on the system structure, or can also be verified by employing the Hamiltonian estimation result to predict the system evolution dynamics and compare the prediction with actual experiment results. Since we are taking Pauli basis, this sparse representation in multi-qubit systems is general enough to allow for multi-axis, complicated and non-classical correlations between different qubits. The qubits number $N$ is not restricted and can be up to $30$ or more in principle.

For the details to perform CS, the specific scenario we are interested in is when $\delta_k < \sqrt{2} - 1$, where according to the results in \cite{PRA-Main,CANDES}, provided that
\begin{equation}
    m \geq c_0 k \cdot\log (\frac{4^N}{k}),
\end{equation}
by solving the following convex optimization problem with a small permissible error $\epsilon>0$,
\begin{equation}\label{op}
    \begin{split}
        \min &\|h\|_1\\
        s.t.\,\|\overline{y}-&\Phi h\|_2\leq \epsilon,
    \end{split}
\end{equation}
we can obtain a good estimation $h^*$ of the true solution to \eqref{mat eq} (denoted as $h_0$), and
it holds with probability $\geq 1-2e^{-c_1m}$ that 
\begin{equation}
    \|h^*-h_0\|_2 \leq \frac{d_1}{\sqrt{k}}\|h_0(k)-h_0\|_1+d_2\epsilon,
\end{equation}
where $h_0(k)$ is the vector obtained by $k$-sparsifying the true Hamiltonian $h$, and $c_0,c_1,d_1,d_2$ are constants independent of $k$. In our paper, we first need to randomly select $m$ pairs $(M_j,\rho_i)$ according to a proper probability distribution, and the associated expectation is denoted as $E(\cdot)$ throughout this paper. Then we need to compute the measurement matrix $\Phi$, substitute $\Phi$ into the optimization problem \eqref{op}, and solve it to obtain the estimate $h^*$. There are mature approaches to solve problem (15) such as the numerical solver ``CVXPY"\cite{cvxpy} (which is the way employed in our simulation).

We highlight that unlike in standard compressed sensing methods, here the elements of our measurement matrix $\Phi$ are not all i.i.d. It can be observed that while the elements within each column of $\Phi$ are i.i.d., the elements in the same row are in fact correlated. Nevertheless, according to the study in \cite{PRA-Main}, a sufficient condition for RIP to hold with probability $1-2e^{-mc_2(\delta+c_3)^2}$ ($c_2,c_3$ are constants) can be derived using measure concentration properties of random matrices as
\begin{equation}\label{cond}
   \delta_k<\delta ,\,\forall \delta\in(\frac{\kappa-1}{\kappa+1},1),\,\forall k\in \mathbb{N}^+,
\end{equation}
where $\kappa$ is the ratio of the maximal singular value to the minimal singular value of $E(\Phi^{\dagger}\Phi)$. (This ratio is also commonly referred to as the \textit{condition number} of the matrix $E(\Phi^{\dagger}\Phi)$.) We can see that $\frac{\kappa-1}{\kappa+1}$ is an upper bound for $\delta_k$.
Since we hope $\delta_k<\sqrt{2}-1$, it would be favorable to have $\frac{\kappa-1}{\kappa+1}<\sqrt{2}-1$, which is equivalent to $\kappa<\sqrt{2}+1$. Although it is only a sufficient condition, it is by far not clear how to restore RIP when this condition is violated.

\subsection{Scale Transformation Method}\label{STM}
\begin{figure}[htbp]
\centerline{\includegraphics[width=0.5\textwidth]{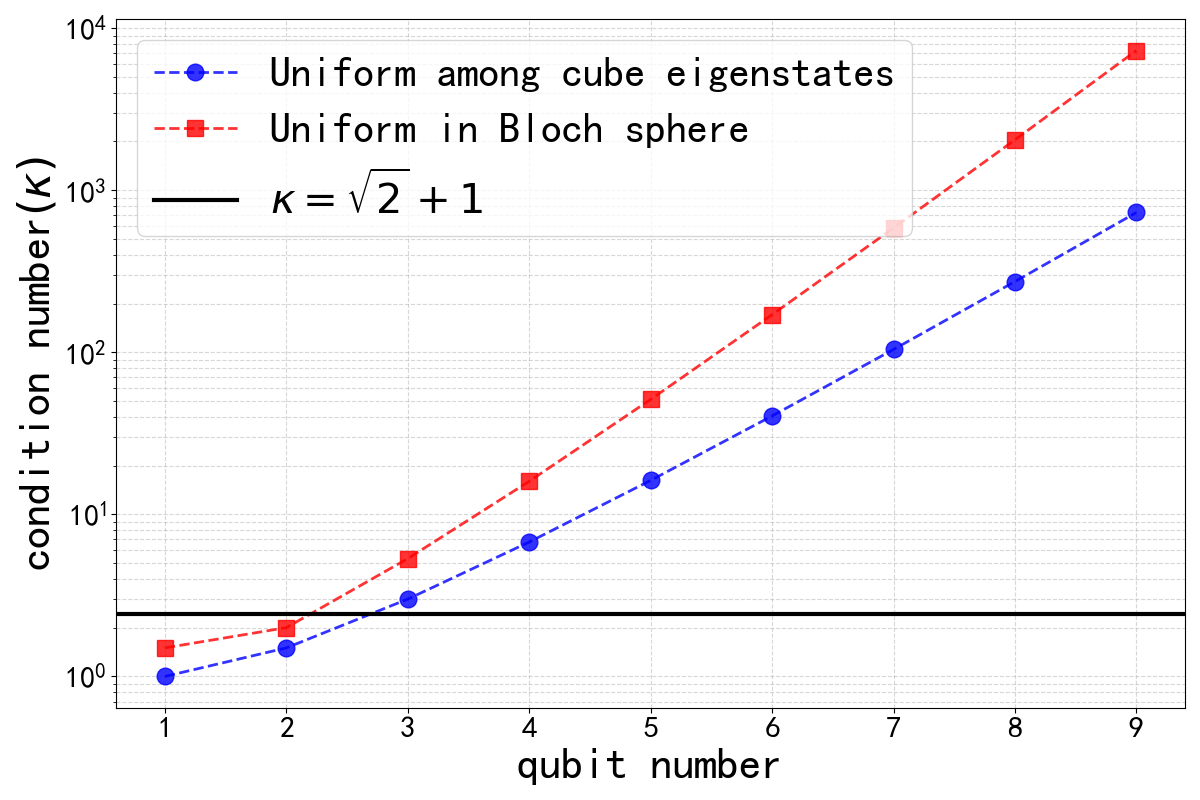}}
\caption{The condition number of $E(\Phi^\dagger \Phi)$ versus the qubit number , with the initial state being tensor product of single qubit each randomly generated from six eigenstates of cube measurement with equal probability (blue dots), or from a uniform probability distribution in Bloch sphere (red squares), and the black line shows the location where condition number is $\sqrt{2}+1$.}
\label{figill}
\end{figure}

Based on the preceding discussion, we have established that to successfully implement the compressed sensing algorithm to estimate a sparse Hamiltonian, the condition $\kappa < \sqrt{2} + 1$ had better be satisfied. However, during our research, we identify a problem we refer to as the ``scalability-induced ill-conditioning problem". This issue manifests as a gradual deterioration of the $\kappa$ value when the number of qubits $N$ increases, such that for multi-qubit systems the RIP often fails. Fig.~\ref{figill} illustrates the trend of the condition number of $E(\Phi^\dagger \Phi)$ as the number of qubits increases under two different initial state distributions, where the ``Uniform among cube eigenstates" means multi-qubit cube state composed of the tensor product of single-qubit states, each randomly selected from $\{\frac{I\pm X}{2},\frac{I\pm Y}{2},\frac{I\pm Z}{2}\}$ with equal probability, and the ``uniform in Bloch sphere" is the tensor product of single qubit $\frac{I+\sin{\theta}\cos{\varphi}X+\sin{\theta}\sin{\varphi}Y+\cos{\theta}Z}{2}$, where $\varphi$ and $\theta$ are uniform distribution in $[0,2\pi)$ and $[0,\pi)$, respectively. It can be observed that, despite the differences in their magnitudes, both sets of data points in Fig.~\ref{figill} exhibit a fast increase as the qubit number increases, and surpass the critical value $\sqrt{2}+1$ for $N\geq 3$ qubits.

To deal with this ill-conditioning problem, we develop a method called ``scale transformation", whose underlying mechanism resembles a form of regularization method. We note that for many common probability distributions (and recall that the expectation on this is denoted as $E(\cdot)$) following which the random matrix $\Phi$ is generated, the matrix $E(\Phi^{\dagger}\Phi)$ is often diagonal, as stated below. 
\begin{proposition}\label{prop1}
   For an $N$-qubit system, if the observable takes any Pauli matrix with equal probability, and the initial state is the tensor product of random single-qubit pure states, where the random Bloch sphere coordinates $(x,y,z)$ of each single-qubit pure state satisfy $E(x)=E(y)=E(z)=E(xy)=E(xz)=E(yz)=0$, then the resulting measurement matrix $\Phi$ satisfies that $E(\Phi^{\dagger}\Phi)$ is diagonal. 
\end{proposition}

Under such circumstance, we can calculate a transformation matrix $\Lambda$ as
\begin{equation}\label{lambda}
    \Lambda_{i,j}=\delta_{ij}\sqrt{E(\Phi^{\dagger}\Phi)_{i,i}},
\end{equation}
where $\delta_{ij}$ is the Kronecker Delta function, and then transfer $\Phi$ to $\tilde{\Phi}=\Phi\Lambda^{-1}$. By introducing a transformed vector $x\equiv \Lambda h$, we obtain an equivalent new linear system of equations
\begin{equation}\label{trans mat}
    \overline{y}=\tilde{\Phi}x,
\end{equation}
where the true solution $x_0$ can be related to the true parameterized Hamiltonian $h_0$ by $h_0=\Lambda^{-1}x_0$. Obviously, the condition number $\kappa$ of $E(\tilde{\Phi}^{\dagger}\tilde{\Phi})$ is 1. Moreover, $x$ remains the same sparsity as $h$. So we can apply the compressed sensing algorithm \ref{CSAL} to the new model \eqref{trans mat} to obtain an estimation $x^*$, and then $h^*=\Lambda^{-1}x^*$ is the final estimation result.

\subsection{Series Expansion with Second-order Correction Method}
To mitigate the error introduced by the first-order approximation of \eqref{ob eq}, here we introduce an optional technique by first treating \eqref{ob eq} as a function of $t$ and expanding it to the second order
\begin{equation}\label{2nd eq}
    y_{ij}=\mathrm{Tr}(M_j\rho_i)+K_1t+K_2t^2+O(K_3t^3),
\end{equation}
where $K_1,K_2,K_3$ are constants for given $i,j$. By measuring several sets of $y_{ij}$ at different evolution times $t_1$, $t_2$, ..., the values of $K_1$ and $K_2$ can be estimated by solving a system of linear equations
\begin{equation}\label{2nd eqs}
    \begin{pmatrix}
        \bar{y}_{ij,t_1}\\
        \bar{y}_{ij,t_2}\\
        \cdots
    \end{pmatrix}=
    \begin{pmatrix}
        t_1 & t_1^2\\
        t_2 & t_2^2\\
        \cdots & \cdots
    \end{pmatrix}
    \begin{pmatrix}
        K_1\\
        K_2
    \end{pmatrix}+
    \begin{pmatrix}
        O(K_3t_1^3)\\
        O(K_3t_2^3)\\
        \cdots
    \end{pmatrix},
\end{equation}
with the third-order term serving as a perturbation term in the equation. Then in \eqref{dot eq}, we modify $\bar{y}_{ij}$ to $\bar{y}_{ij}-K_2 t^2$, thereby reducing the error introduced by the first-order approximation. Assume that the original estimation method consumes $\bar{N}$ state copies for each \eqref{dot eq}, and that $\bar{K}$ different $t_i$ are used in this second-order correction method (in our simulation we take $\bar{K}=2$) to correct a single \eqref{dot eq}, then it should take $\bar{N}/\bar{K}$ measurement shots for each evolution time $t_i$ in this correction method, in order to remain the total resource consumption (measurement shots) unchanged. Under this circumstance, the optional second-order correction method can often reduce the approximation error in data, according to our simulation experience.

\subsection{Error Analysis}\label{EA}
The errors in our model primarily mainly arise from the following aspects. 
Firstly, there are errors inherent to the model itself. Our model primarily focuses on time-independent closed systems. Although it can be straightforwardly extended to time-independent open systems (to be elaborated in future work), non-trivial modeling errors persist when dealing with actual time-dependent closed systems and time-dependent open systems. Additionally, the accuracy of prior information regarding sparsity also affects the precision of the estimation results.

The first-order approximation error introduced by \eqref{ob eq} is of the order $O(S^2t^2)$. Beyond the first-order approximation error, the quantum expectation itself inherently incurs errors, as its exact value cannot be obtained from finite measurement shots in practice. The error $err$ follows a normal distribution $err\sim \mathcal{N}(0,\frac{\sigma^2}{n})$, where $n$ is the number of copies of the quantum state, and $\sigma^2=\langle M_j^2\rangle_{p_i}-\langle M_j\rangle_{p_i}^2$. Since the operational duration of quantum devices is limited, the number of copies for each state cannot be increased indefinitely. Consequently, we must sometimes tolerate the errors arising from estimating quantum expectation.

The scale transformation method also introduces errors. The transformation matrix $\Lambda$ is diagonal, and the variation in the magnitudes of its diagonal entries leads to inconsistent amplification (or attenuation) of the components of the estimated Hamiltonian. When the non-sparse terms in the actual Hamiltonian are not strictly zero, some weak multi-body terms may be amplified, thereby affecting the sparsity of the transformed estimate and resulting in estimation errors.

Due to the complexity of quantum systems, there are numerous sources of error. Given space limitations, this paper provides a brief overview of several main sources of error, while a more detailed analysis is not discussed here.
\subsection{Algorithm}\label{Alg}
Here we list the general procedures to apply our CS reconstruction of a sparse Hamiltonian in Algorithm \ref{al}. The main version of the algorithm is without the second-order correction method, which is presented as an optional substitution in Step 2.
\begin{algorithm}
\caption{CS algorithm of Hamiltonian estimation}\label{alg:cap}
\renewcommand{\algorithmicrequire}{\textbf{Input:}}
\renewcommand{\algorithmicensure}{\textbf{Output:}}

\begin{algorithmic}\label{al}
\REQUIRE Qubit number $N$, sparsity $k$, state copy number $N_S$ for each equation, evolution time $t$, randomly selected (satisfying the requirements in Proposition \ref{prop1}) $m$ pairs of observable $M_j$ and initial state $\rho_i$.
\ENSURE An estimated Hamiltonian $\bar{H}$.
\STATE \textbf{Step 1:} Compute the measurement matrix $\Phi$ 
    $$\phi_{ij,\alpha}=\text{i}t(\mathrm{Tr}(\Gamma_{\alpha}[M_j,\rho_i])$$
for each of the $m$ of $(M_j,\rho_i)$ pairs as well as those $\alpha$ with which we known a prior that $h_\alpha$ is non-zero. 
\STATE \textbf{Step 2:} Measure $M_j$ on $N_S$ copies of $\rho_i(t)$ and take the outcomes' average as $y_{ij}$. Compute $\bar y$ from
$$\bar{y}_{ij}=y_{ij}-\mathrm{Tr}(M_j\rho_i).$$
\STATE \textbf{Step 3:} Compute the transformation matrix $\Lambda$ from
    $$\Lambda_{i,j}=\delta_{ij}\sqrt{E(\Phi^{\dagger}\Phi)_{i,i}},$$
    and let $\tilde{\Phi}=\Phi\Lambda^{-1}.$
\STATE \textbf{Step 4:} Solve the following convex optimization problem to obtain a solution $x^*$,
    \begin{equation}
        \begin{split}
        \min &\|x\|_1\\
        s.t.\,\|\bar{y}-&\tilde{\Phi} x\|_2\leq \epsilon,
        \end{split}
    \end{equation}
    where $\epsilon>0$ is a given small constant.
\STATE \textbf{Step 5:} Calculate the estimation as $h^*=\Lambda^{-1}x^*$ and
$\bar{H}$=$\sum_{\alpha}h_{\alpha}^*\Gamma_{\alpha}$.
\RETURN $\bar{H}$.
\end{algorithmic}
\end{algorithm}

\subsection{An Example for Further Application -- Optimal System Control}
In this manuscript, since we have already explained the Hamiltonian estimation approach based on CS, we present the task of optimal system control in this subsection, as an example of many subsequent applications based on the estimation result of the system Hamiltonian. But before this, we first briefly introduce Dynamical Decoupling (DD). DD is a powerful quantum control 
technique used to suppress noise or decoherence in quantum systems by applying a sequence of periodic or carefully timed control pulses\cite{DD1,DD2}. These pulses effectively average out unwanted environmental interactions, thereby preserving the coherence of quantum states. Originally developed in nuclear magnetic resonance, DD has become a fundamental tool in quantum information processing\cite{DD3}. By appropriately designing the pulse sequences, one can selectively decouple the system from specific noise sources, making DD particularly valuable for quantum sensing, quantum computing, and precision metrology applications.

The pulse sequence commonly used by DD on a single qubit is a train of $\pi$ pulses, each around a Pauli operator. In a multi-qubit system, for simplicity we only consider local control enforced individually on each qubit, instead of global control. Their effect can be viewed as multiplying the Hamiltonian by a switch function, i.e.,
\begin{equation}
    \widetilde{H}(t)=y(t)H(t),
\end{equation}
where $\widetilde{H}(t)$ is the Hamiltonian obtained after moving to the interaction picture w.r.t. the control Hamiltonian, $y(t)$ is the switch function which takes value in $\{-1,1\}$. Starting from $y(0)=1$, the sign of $y(t)$ flips whenever a pulse array anti-commuting with $H(t)$ is enforced. According to the time-dependent Schrödinger equation, the evolution operator $\widetilde{U}$ of the system after DD takes the following form
\begin{equation}\label{evolve}
    \widetilde{U}=\mathcal{T}\exp\left(-\text{i}\int_0^t y(t')H(t')\,dt'\right),
\end{equation}
where $\mathcal{T}$ is the time-ordering superoperator. Since we are focusing on time-independent Hamiltonian $H(t)\equiv H$, we can further write \eqref{evolve} as
\begin{equation}
\widetilde{U}=\exp\left(-\text{i}H\int_0^t y(t')dt'\right).
\end{equation}
By carefully designing the pulse sequence, one can achieve $\int_0^t y(t')dt'=0$, which leads to $\widetilde{U}=I$. This net effect is desirable in many tasks, as it eliminates or suppresses unwanted Hamiltonian. Hence, when a quantum system needs to be decoupled from environment Hamiltonian, DD can be an effective candidate technique which protects the quantum information from decoherence by making no measurements.

Now we present how to apply our method in the field of optimal control, for instance, in state preservation. For state preservation tasks, a common approach involves DD sequences to suppress the system Hamiltonian and extend coherence time, thereby achieving state preservation. However, traditional DD sequence design, such as the generic XY4 sequence\cite{DD2}, does not incorporate information about the system Hamiltonian. In contrast, employing the Hamiltonian information when designing specialized DD sequences can in principle wipe out the unwanted Hamiltonian cleaner than universal DD sequences. To this end, we can formulate the following optimal control model
\begin{equation}
    \begin{split}
        \min_{DD\in O} &\|\rho_t-\rho_0\|_{Fro},\\
        s.t.\,\widetilde{U}=&\exp\left(-\text{i}H\int_0^t y(t')dt'\right),\\
        \rho_t=&\widetilde{U}\rho_0\widetilde{U}^{\dagger},
    \end{split}
\end{equation}
where $O$ is the set of all possible DD sequences, $H$ is the original Hamiltonian of the system, $\rho_0$ is a given initial state, $\rho_t$ is the final state after the initial state has evolved for time $t$, $\|\cdot\|_{Fro}$ is the Frobenius norm. According to our prior knowledge, we assume that the actual number of many-body terms in the Hamiltonian does not grow exponentially with the number of qubits. Therefore, our compressed sensing algorithm can still operate effectively, making this optimal control model applicable to systems with up to 30 or even more qubits.

\section{Numerical examples}\label{NE}

In our simulation, the test states are pure states. For the $i$-th qubit, its initial state is selected according to 
\begin{equation}
    \rho^i=\frac{1}{2}(I+\frac{x_iX+y_iY+z_iZ}{\sqrt{x_i^2+y_i^2+z_i^2}}),
\end{equation}
where $x_i,y_i,z_i$ are independently sampled from normal distribution, and each qubit is independently generated in this way, to form a random test state $\rho=\bigotimes_{i=1}^N\rho^i$, where $\bigotimes$ means tensor product.
The observable satisfies
\begin{equation}
    M_j=\bigotimes_{l=1}^N(\delta_{la}(\delta_{1b}X+\delta_{2b}Y+\delta_{3b}Z)+(1-\delta_{la})I),
\end{equation}
where $a$ is chosen uniformly at random from 1 to $N$, and $b$ is chosen uniformly at random from 1 to 3.

\begin{figure}[htbp]
\centerline{\includegraphics[width=0.5\textwidth]{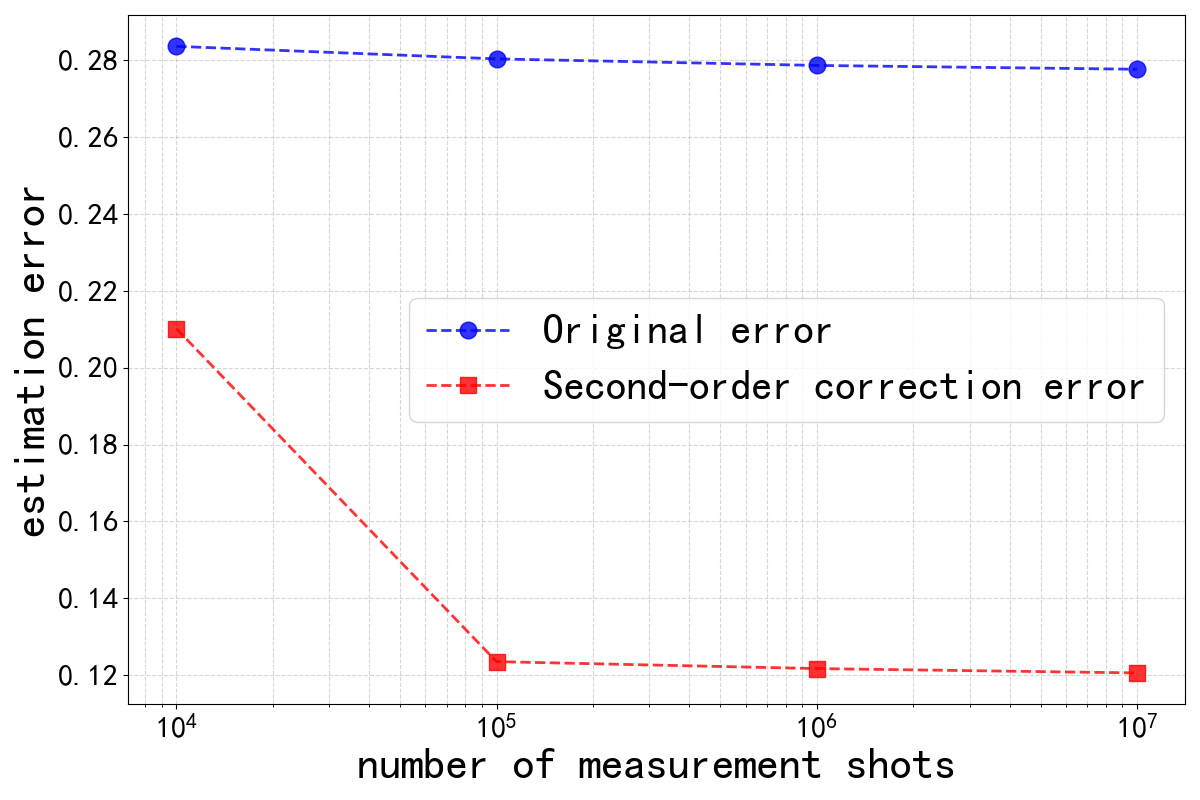}}
\caption{Simulated estimation errors for a 6-qubit Hamiltonian}
\label{fig1}
\end{figure}

We selected a 6-qubit time-dependent open system for simulation, and its Hamiltonian has the form
\begin{equation}\label{hform}
     H(t)=\sum_{i=1}^{6}h_i(1+b_i(t))\sigma_{i_1}^i+\sum_{j=1}^{5}h'_j
     (1+b'_j(t))\sigma_{j_1}^j\sigma_{j_{2}}^{j+1},
\end{equation}
where $\sigma_{\alpha}^i$ means the Pauli operator $\sigma_{\alpha}$ on the $i$-th qubit and identity on the other qubits, $b_i(t),b'_j(t)$ are the perturbation noise arising from coupling to the bath, which we model as independent Gaussian stochastic processes with the mean $0$ and variance 0.01. In our test, the Hamiltonian has a sparsity of 26, with the values of the sparse terms independently drawn from a standard normal distribution. A total of 80 pairs $(M_j,\rho_i)$ were randomly and independently selected, yielding 80 corresponding measurement equations. The estimation errors of the Hamiltonian obtained by the original method and by employing the second-order correction are shown in Fig.~\ref{fig1}, and the error is defined by $error\equiv\frac{\|h^*-h_0\|_1}{\|h_0\|_1}$, where $h_0\equiv (h_i^T, (h_j')^T)^T$ is the actual mean of the unknown Hamiltonian parameters, and $h^*$ the corresponding estimation. The horizontal axis represents the number of measurement shots for each \eqref{dot eq}.
It can be observed that the introduction of the second-order correction method significantly reduces the estimation error. As the number of measurement shots increases, the estimation error steadily decreases, indicating that the compressed sensing algorithm is feasible for Hamiltonian estimation. However, the rate of decrease gradually slows down, and even infinite measurement shots cannot reduce the estimation error to zero, because statistical error is not the only source of error.

Considering only time-independent closed systems, the application range of our CS algorithm can be extended to more than 30 qubits. In the simulation, the form of Hamiltonian is similar to \eqref{hform}, while only the non-random terms are retained and the many-body terms involve more than just neighboring qubits as
\begin{equation}\label{hform30}
     H(t)=\sum_{i=1}^{30}h_i\sigma_{i_1}^i+\sum_{1\leq j<l\leq 30}h_{jl}\sigma_{j_1}^j\sigma_{l_{1}}^{l}.
\end{equation}
We randomly choose 700 pairs $(M_j,\rho_i)$ which yields 700 equations, and the Hamiltonian has a sparsity of 100, while in the absence of location information about the sparse terms, conventional methods would need to consider 4005 parameters, generating at least 4005 equations.

\begin{figure}[htbp]
\centerline{\includegraphics[width=0.5\textwidth]{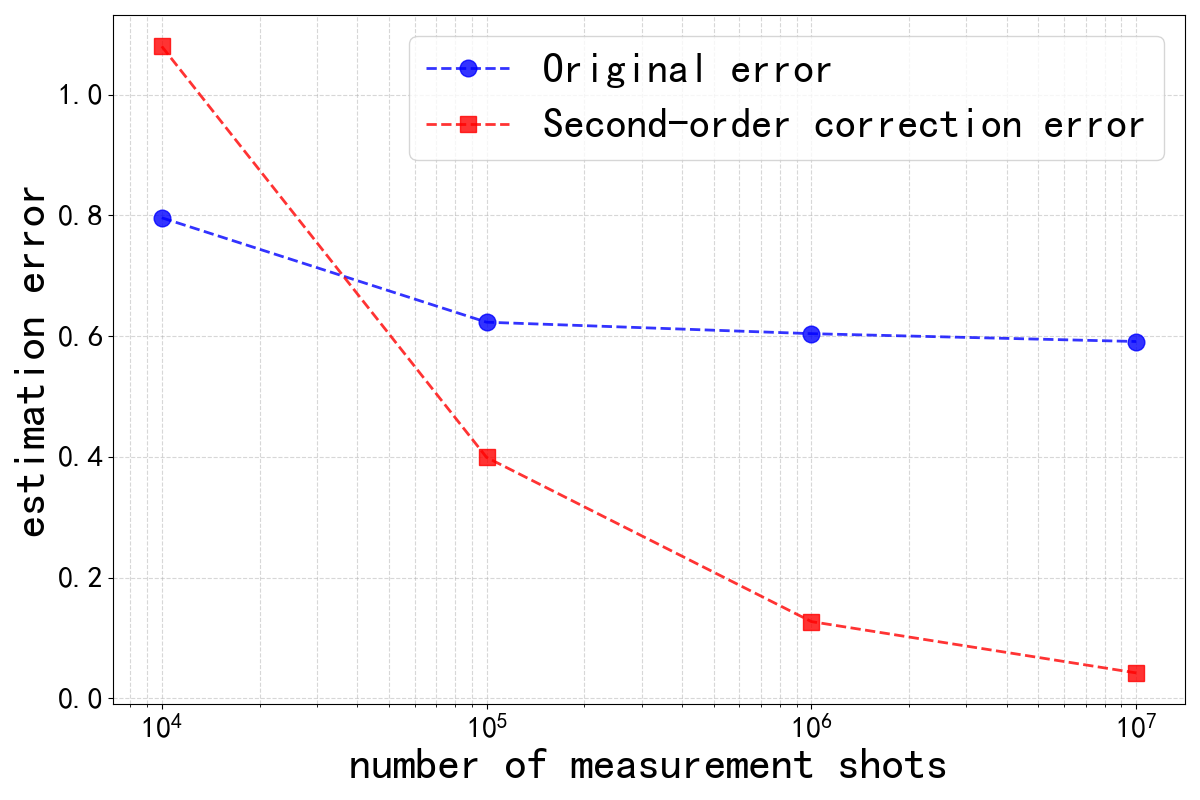}}
\caption{Simulated estimation errors for a 30-qubit Hamiltonian}
\label{fig30b}
\end{figure}

Fig.~\ref{fig30b} shows the estimation error of the CS as a function of the number of measurement shots in the system \eqref{hform30}. It can be seen that our method, even with the number of qubits reaching 30, can still provide estimates within a reasonable error range when the number of measurement shots is large enough by utilizing the second-order correction method, indicating that this method has the potential to be extended to systems with a larger number of qubits. Subsequently, the reason for the inferior performance of the original method is that, as the number of qubits increases, the model error becomes increasingly dominated by the first-order approximation error, thus limiting the effectiveness of increasing the measurement shots. The exceptional data of $10^4$ measurement shots, where the original method outperforms the second-order correction method, is because the statistical error outweighs the first-order approximation error, and thus the second-order correction method gains little, meanwhile suffering from a condition number a little larger than $1$ when solving \eqref{2nd eqs}.

\section{Conclusion}\label{CL}

We have proposed a method for Hamiltonian estimation using compressed sensing, which effectively reduces the number of different measurements required in an $N$-qubit system from $\Omega(4^N)$ to $O(k\log (4^N/k))$, where $k$ is the sparsity of the Hamiltonian parameterized in a known basis. We identified a ``scalability-induced ill-conditioning problem", meaning that as the qubit number increases, the condition number of the expectation of $\Phi^T \Phi$ (where $\Phi$ is the sensing matrix) often increases such that the RIP condition for compressed sensing cannot be directly satisfied by most common random sampling rules. To solve this problem, we introduced a ``scale transformation" method that ensures the RIP even for increasing $N$. We also proposed a second-order correction method to reduce the first-order approximation error in measurement data. Our simulation results demonstrate that the proposed method maintains applicable and effective even in time-varying multi-qubit systems. In the future, we aim to verify the performance of this method on experimental systems such as superconducting qubits.

\appendix
In Section \ref{STM}, we propose a method called ``scale transformation'' to address the issue of scalability-induced ill-conditioning. The foundation of this method is the scenario where $E(\Phi^{\dagger}\Phi )$ is diagonal. To this end, we provide a sufficient condition (Proposition \ref{prop1}) such that when states and observables are randomly selected according to its specification, it holds that $E(\Phi^{\dagger}\Phi )$ is diagonal.
Here we prove Proposition \ref{prop1} in the $N$-qubit system. Since we have
\begin{equation}
    \Phi=\begin{pmatrix}
        \phi_1
        ,\phi_2
        ,\cdots
        ,\phi_m
    \end{pmatrix}^T,
\end{equation}
where each row $\phi_i^T$ of $\Phi$ is i.i.d., we have
\begin{equation}
    \begin{split}
        E(\Phi^{\dagger}\Phi )&=E\left(\sum_{i=1}^m(\phi_i^T)^{\dagger}\phi_i^T\right)\\
        &=mE\left((\phi_q^T)^{\dagger}\phi_q^T\right)
    \end{split}    
\end{equation}
for any $1\leq q\leq m$. Then the element at position $(\alpha,\beta)$ of $E(\Phi^T \Phi)$ is
\begin{equation}\label{expec eq}
    E(\Phi^{\dagger}\Phi )_{\alpha,\beta}=mE(\mathrm{Tr}(\Gamma_{\alpha}[M_j,\rho_i])^{\dagger}\cdot \mathrm{Tr}(\Gamma_{\beta}[M_j,\rho_i]))
\end{equation}
for any proper $i$ and $j$. Denote
\begin{equation}
    \Gamma_{\alpha}=\bigotimes_{i=1}^N\sigma_{\alpha_i},
\end{equation}
where $\sum_{i=0}^{N-1}\alpha_{N-i}4^i=\alpha$.

As $M_j$ takes $\sigma_b$ ($1\leq b\leq 3$) matrix on the $a$-th qubit and identity on all the other qubits, we denote
\begin{equation}
    M_j=\sigma_b^a.
\end{equation}
We also denote
\begin{equation}
    \rho_i=\bigotimes_{k=1}^N\rho_i^k,\,\rho_i^k=\frac{1}{2}(I+x\sigma_1+y\sigma_2+z\sigma_3),
\end{equation}
and remember that we have
\begin{equation}
    E(x)=E(y)=E(z)=E(xy)=E(xz)=E(yz)=0.
\end{equation}
We focus on the case $\alpha \neq \beta$, where there exists at least one $i_0 \in \{1,2,\cdots,N\}$ such that $\alpha_{i_0} \neq \beta_{i_0}$. For $M_j$, without loss of generality we assume $a=1$. Hence we have
\begin{equation}
    \begin{split}       
    &\mathrm{Tr}(\Gamma_{\alpha}[M_j,\rho_i])\\&=\mathrm{Tr}(\sigma_{\beta_1}[\sigma_b,\rho_i^1]\otimes\sigma_{\beta_2}\rho_i^2\otimes\cdots\otimes\sigma_{\beta_N}\rho_i^N)\\    &=\mathrm{Tr}(\sigma_{\beta_1}[\sigma_b,\rho_i^1])\mathrm{Tr}(\sigma_{\beta_2}\rho_i^2)\cdots \mathrm{Tr}(\sigma_{\beta_N}\rho_i^N).
    \end{split}        
\end{equation}
Then
\begin{equation}
    \begin{split}
        &E(\mathrm{Tr}(\Gamma_{\alpha}[M_j,\rho_i])^{\dagger}\cdot \mathrm{Tr}(\Gamma_{\beta}[M_j,\rho_i]))\\
        &=E(\mathrm{Tr}(\mathrm{Tr}(\sigma_{\alpha_1}[\sigma_b,\rho_i^1])\mathrm{Tr}(\sigma_{\alpha_2}\rho_i^2)\cdots \mathrm{Tr}(\sigma_{\alpha_N}\rho_i^N))\\
        &\,\,\,\,\,\,\,\, \mathrm{Tr}(\sigma_{\beta_1}[\sigma_b,\rho_i^1])\mathrm{Tr}(\sigma_{\beta_2}\rho_i^2)\cdots \mathrm{Tr}(\sigma_{\beta_N}\rho_i^N))\\
        &=E(\mathrm{Tr}(\sigma_{\alpha_1}[\sigma_b,\rho_i^1]\mathrm{Tr}(\sigma_{\beta_1}[\sigma_b,\rho_i^1]))E(\mathrm{Tr}(\sigma_{\alpha_2}\rho_i^2)\mathrm{Tr}(\sigma_{\beta_2}\rho_i^2))\\
        &\,\,\,\,\,\,\,\,\cdots E(\mathrm{Tr}(\sigma_{\alpha_N}\rho_i^N)\mathrm{Tr}(\sigma_{\beta_N}\rho_N^2)).
    \end{split}
\end{equation}
If $i_0=1$, since $b$ takes value in $\{1,2,3\}$, we know $[\sigma_b, \rho_i^1]=[\sigma_b,I+x\sigma_1+y\sigma_2+z\sigma_3]/2=\text{i}\gamma_1\sigma_{b_1}-\text{i}\gamma_2\sigma_{b_2}$, where $b_1=(b+1\ mod\ 3)+1$, $b_2=(b\ mod\ 3)+1$, and $\gamma_1$ and $\gamma_2$ are two different variables of $\{x,y,z\}$. Hence,
\begin{equation}
    \begin{split}
    & E(\mathrm{Tr}(\left[\sigma_{\alpha_1},\sigma_b\right]\rho_i^1)\mathrm{Tr}(\left[\sigma_{\beta_1},\sigma_b\right]\rho_i^1))\\
    &\,\,\,\,\,\,\,\,\,\in\{0,E(4xy) ,E(4yz),E(4xz)\} =\{0\}.
    \end{split}
\end{equation}
When $i_0>1$,
\begin{equation}
    \begin{split}
        &E(\mathrm{Tr}(\sigma_{\alpha_i}\rho_i^k)\mathrm{Tr}(\sigma_{\beta_i}\rho_i^k))\\
        &\in \{E(x),E(y),E(z),E(xy) ,E(yz),E(xz)\}=\{0\}.
    \end{split}   
\end{equation}
Hence when $\alpha\neq \beta$, we always have
\begin{equation}\label{zeroeq}
    E(\mathrm{Tr}(\left[\Gamma_{\alpha},M_j\right]\rho_i)^{\dagger} \mathrm{Tr}(\left[\Gamma_{\beta},M_j\right]\rho_i))=0.
\end{equation}
Using \eqref{expec eq}, we know $E(\Phi^{\dagger}\Phi )$ is diagnal.

\bibliographystyle{ieeetr}         
\bibliography{trueref} 

\end{document}